\documentclass[12pt,a4paper]{article}
\usepackage{amsmath}
\usepackage{graphics}

\newcommand{\hepth}[1]{{\tt hep-th/#1}}

\newcommand{\nn}{\nonumber}

\newcommand{\p}{\vspace{6pt}\noindent}
\advance\textheight by \topskip
\makeatletter

\@addtoreset{equation}{section}
\def\section{\@startsection {section}{1}{\z@}{-8.5ex plus -1ex minus
 -.2ex}{3.3ex plus .2ex}{\large\bf}}
\def\subsection{\@startsection{subsection}{2}{\z@}{-3.25ex plus
 -1ex minus -.2ex}{1.5ex plus .2ex}{\bf}}
\def\subsubsection{\@startsection{subsubsection}{3}{\z@}{-3.25ex plus%
 -1ex minus -.2ex}{1.5ex plus .2ex}{\sl}}

\begin{document}
\begin{titlepage}
\vspace*{-2cm}
\begin{flushright}

 hep-th/0407199 \\

\end{flushright}

\vspace{0.3cm}

\begin{center}
{\Large {\bf Aspects of sine-Gordon solitons, defects and gates. }}\\
\vspace{1cm} {\large  E.\ Corrigan and
C.\ Zambon}\\
\vspace{0.3cm} {$\ $\em\it Department of
Mathematics\ \footnote{\noindent E-mails: {\tt ec9@york.ac.uk;\ cz106@york.ac.uk}}  \\
University of York\\
York YO10 5DD, U.K. }\\ \vspace{1cm} {\bf{ABSTRACT}}
\end{center}

\begin{quote}
It was recently noted how the classical sine-Gordon theory can support
discontinuities, or `defects',
and yet maintain integrability by preserving sufficiently many conservation laws.
Since soliton number is not preserved by a defect, a possible application to the
construction of logical gates is suggested.
\end{quote}

\vfill
\end{titlepage}

\section{Introduction}
\label{s:intro} In two recent articles \cite{bczlandau, bczjhep}, based principally on the
affine Toda series of integrable models, of which the sinh-Gordon
(or sine-Gordon) model is the simplest example,  
it was pointed
  out that integrable field theories in
$1+1$ dimensions allow internal boundary conditions which preserve
integrability.

\p
Typically,
at an internal boundary the classical field will have a
discontinuity, hence the name `defect', yet energy (and momentum)
are conserved after they have been suitably modified to take
into account the energy (and momentum) stored in the defect.
Actually, this is already surprising given that translation invariance is
destroyed by placing a defect in a particular location. Moreover, integrability is
maintained in the sense that it is
possible to construct a Lax pair incorporating a defect which will
guarantee (and indeed generate explicitly) an infinity of other
independent conserved quantities.  The properties of a single
defect can be repeated since any
number of defects may be placed along the $x$-axis - each bringing
an additional parameter to the model. In \cite{bczlandau} it was
pointed out how a  single soliton solution is affected when the
soliton encounters a defect. In this letter it is intended to
expand upon this observation and point out a possible use.
Naturally, it must be said at the outset that this work is entirely
theoretical and speculative since, although it appears to be the case
that sine-Gordon or other solitons can appear within special
systems, for example Josephson junctions, polymers,  or liquid
crystals (see for example  \cite{Ustinov98, Zubova01, Khoo93} and references therein),
it is not yet known how the specific defect
introduced and described mathematically in \cite{bczlandau} might be
realised in an a genuine physical system.

\p
Solitons have remarkable properties (there are many reviews, but see for example
 \cite{Scott73, Lamb80}) and many uses have been proposed in different contexts. In the
arena of information theory or computation several ideas suggest themselves.
For example, a soliton is stable and therefore might be used reliably to transport a bit
of information. The sine-Gordon equation, being relativistic, requires that solitons
have a maximum speed above which they cannot travel, and a train of separated
solitons all moving
close to that speed might be used reliably and efficiently to convey data.
More remarkably, sine-Gordon
solitons have a scattering property, in the sense that a fast soliton behind a slow
soliton will inevitably overtake it but neither will lose their integrity. All that happens
is a relative delay (a shift forwards for the faster soliton and a shift backwards for
the slower one). One could imagine making use of this property also in a scenario where
data is transported relatively
slowly and a fast soliton might be used to overtake the data and signal to the receiver
information concerning the data
(that it should be ignored, for example). Besides solitons, there are also anti-solitons
and these scatter maintaining their integrity not only with other anti-solitons but also
with solitons. In particular, a soliton will not annihilate an antisoliton.
A soliton and an anti-soliton
can make a permanently bound state (a `breather') with
centre of mass energy $0<E<2M_s$, where $M_s$
is the mass of the soliton/anti-soliton.

\p
One idea which has appeared in the literature, concerning the use of solitons to perform
logical operations, seems  radical
because it is envisaged that the whole computation takes place using the dynamics of
certain solitons (not necessarily those associated with the sine-Gordon model),
dispensing entirely with standard gates \cite{Jakubowski01}.
The purpose of this note is to point out that conventional computing might
be carried out, albeit theoretically, by using solitons and anti-solitons to carry
data, capitalising on the properties of defects to construct standard logic gates.

\section{Defects in the sine-Gordon model}
\label{s:fsf}

First, the defect idea introduced in \cite{bczlandau} will be reviewed briefly,
following the
notation and conventions established previously, and it is enough in the first
instance to consider a single defect located at $x=0$.

\p
For convenience, the field in the region $x>0$ will be denoted $\phi_2$
and the field in the region $x<0$ will be denoted $\phi_1$. Then, the field
equations in the two regions together
with the defect conditions at $x=0$ are:
\begin{eqnarray}\label{SGdefect}
&&x<0:\quad\partial^2\phi_1=-\sin\phi_1,\quad\\
&&x>0:\quad\partial^2\phi_2=-\sin\phi_2,\\
&&x=0:\quad \partial_x\phi_1-\partial_t\phi_2=
-\sigma\sin\left(\frac{\phi_1+\phi_2}{2}\right)-\frac{1}{\sigma}
\sin\left(\frac{\phi_1-\phi_2}{2}\right)\nn\\
&&\phantom{x=0:}\quad\ \partial_x\phi_2-\partial_t\phi_1=
\phantom{-}\sigma\sin\left(\frac{\phi_1+\phi_2}{2}\right)-\frac{1}{\sigma}
\sin\left(\frac{\phi_1-\phi_2}{2}\right)\label{condition}.
\end {eqnarray}
As pointed out in \cite{bczlandau}, these equations follow from a simple
Lagrangian description of the sine-Gordon model with a defect though there is no need to
review that aspect here. The form of the defect condition \eqref{condition}
is dictated by a desire to
maintain integrability and it has been remarked already
\cite{bczlandau} how similar it is to a B\"acklund
transformation \cite{Back1882}. Usually, a B\"acklund transformation
relates two different solutions to a nonlinear field equation (or solutions
to two different field equations) defined in a common domain. However,
in the present context the two spatial derivatives are frozen at the
location of the defect.
The parameter $\sigma$ is free and associated with the defect.
If $\sigma$ is set to zero, the
two fields on either side of $x=0$ are forced to have the same
value at $x=0$ and the defect disappears. Generally, for  other choices of
$\sigma$ there will a discontinuity since
$\phi_1(0,t)\ne \phi_2(0,t)$. If there are several defects then
each will introduce its own free parameter.

\p
An important feature of a B\"acklund transformation is its ability
to generate (or remove) solitons and in fact this is one method of constructing
multi-soliton solutions by repeated application - see, for example, \cite{Scott73,Lamb80}.
The question arises
concerning the extent to which this property survives when the $x$-derivatives
are frozen, as they are in the above defect condition. If it does, then a
defect has the potential to act as a filter, or `gate', at least for suitable choices
of the parameter $\sigma$.

\p
Consider first a single soliton approaching a defect at $x=0$ from the right ($x>0$).
A convenient expression for the soliton solution in the two regions has the form
(see for example \cite{Hiro72,Hiro80})
\begin{equation}\label{soliton}
e^{i\phi_a/2}=\frac{1+E_a}{1-E_a}, \quad E_a=e^{\alpha_a x
+\beta_a t + \gamma_a}, \quad \alpha_a^2-\beta_a^2=1,\quad a=1,2,
\end{equation}
with $\alpha$ and $\beta$ real, and Im$\gamma =i\pi/2$.
In order to be able to satisfy the conditions \eqref{condition} the
time dependence must match in the two domains (implying $\beta_1=\beta_2$)
and the constants $\gamma_1,\ \gamma_2$ are related by
\begin{equation}\label{delay}
\gamma_1=\gamma_2-\ln\left(\frac{e^\theta +\sigma}{e^\theta -\sigma}\right),
\end{equation}
where it is convenient to define $\alpha_1=\alpha_2=\cosh\theta$
and $\beta_1=\beta_2=\sinh\theta$ (i.e. the soliton velocity is $-\tanh\theta$).
In other words, one effect of the defect is to
delay or advance the soliton as it passes through.

\p
Suppose $\sigma$ is chosen to be positive with $\sigma>1$,
 then there are several interesting features to observe.
\p

\noindent (a) The incoming soliton solution satisfies $e^\theta >\sigma$.
In this case, the soliton is delayed,
though by less the faster it goes; solitons at their limiting speed
($\theta\rightarrow\infty$) are not delayed at all.

\noindent (b) The incoming soliton satisfies $e^\theta =\sigma$. In this
case, the soliton is infinitely
delayed - or swallowed - by the defect. This feature was already pointed
out in \cite{bczlandau}.

\noindent (c) The incoming soliton satisfies $e^\theta <\sigma$. In this case,
the delay acquires
an imaginary part $i\pi$, indicating that the character of the solution $\phi_1$
has changed. In fact,
if $\phi_2$ is a soliton then $\phi_1$ is an anti-soliton, or vice-versa.

\noindent (d) A soliton travelling in the opposite direction ($\theta$ replaced
by $-\theta$) will
not be swallowed by the same defect as at (b). In this case, a fast soliton
will be delayed
and converted to an anti-soliton; with $\sigma >1$ it will never be absorbed.

\p
If $\sigma <1$ the story is similar except the roles of the soliton and
anti-soliton interchange.

\p
The properties (b) and (c) are surprising, especially if one is used to the
idea of topological charge
- or soliton number - being conserved. On the other hand, once there is a defect,
there is no longer
any reason to expect soliton number to be preserved. Indeed, integrating the density for
topological charge gives
\begin{equation}
Q=\int_{-\infty}^0dx \, \partial_x\phi_1+\int^{\infty}_0 dx\,  \partial_x\phi_2
=\phi_2(\infty,t) -\phi_1(-\infty,t) + \phi_1(0,t) -\phi_2(0,t),
\end{equation}
and it becomes clear the difference $\phi_1
-\phi_2$ measures the strength of the defect at $x=0$. Effectively, in cases (b) or (c),
respectively, the defect is storing one or two units of topological charge.
These two phenomena also fit well
with the traditional uses of B\"acklund transformations.

\p
The different effects of the defect on
solitons moving in opposite directions is reflected in an intriguing feature of
\eqref{condition} with respect to its behaviour under time-reversal. In each of the
bulk regions the sine-Gordon equations are invariant separately under the transformations,
$t\rightarrow -t$ and $\phi_a\rightarrow -\phi_a,\ a=1,2$. In contrast,
the defect condition is invariant
only under the combinations of any pair of these, for example $t\rightarrow -t$ and
$\phi_1\rightarrow -\phi_1$ or $\phi_2\rightarrow -\phi_2$, together with
$\sigma\rightarrow 1/\sigma$. For a given choice of $\sigma$ this implies the model loses
its time-reversal invariance, meaning there is no time-reversed process to (b) in which,
for example, a defect might emit a soliton.

\p
If several solitons approach a defect then they interact with it independently of
one another, each being delayed. This is not difficult to check using an
explicit two soliton solution of the form \cite{Hiro72}
\begin{eqnarray}\label{twosoliton}
&&e^{i\phi/2}=\ln\left(\tau_0/\tau_1\right), \quad \tau_p=1+(-)^p (E^{(1 )}+  E^{(2)})
+A_{12}E^{(1)}E^{(2)}, \ p=0,1\nonumber\\
&& A_{12}=-\tanh^2\left(\frac{\theta_1-\theta_2}{2}\nonumber\right)\\
&&E^{(a)}=e^{\alpha^{(a)} x +\beta ^{(a)} t +\gamma ^{(a)}},\ a=1,2
\end{eqnarray}
in each of the two regions, and imposing the boundary condition \eqref{condition}.
The defect condition requires
\begin{eqnarray}\label{twodelay}
\gamma^{(1)}_1=\gamma^{(1)}_2 -\ln\left(\frac{e^{\theta_1} +\sigma}{e^{\theta_1}
-\sigma}\right)\\
\gamma^{(2)}_1=\gamma^{(2)}_2 -\ln\left(\frac{e^{\theta_2} +\sigma}{e^{\theta_2}
-\sigma}\right)
\end{eqnarray}
Note, by adding $i\pi$ to any of the constants $\gamma^{(a)}$, either one or other, or
both, soliton components can be converted to anti-solitons.

\p
An interesting and important
point to note
is that the defect can absorb at most one soliton (or anti-soliton),  given
a suitable $\sigma$, but not both because a genuine two soliton
solution requires $\theta_1 \ne \theta_2$. On the other hand, neither, or  one,
or both may be
converted to a soliton of the opposite character according to the relative magnitudes
of $e^{\theta_1},\ e^{\theta_2}$ and $\sigma$. These observations appear to suggest that
a defect might be used to model logic gates.

\section{The defect as a logical gate}

In this section it will be supposed that $\sigma > 1$ and solitons approach a defect
from the right ($x>0$).

\p
Adopting the convention that a soliton represents {\it true} or `1', and an anti-soliton
{\it false} or `0', the simplest gate to model is NOT since it is enough that the soliton
(or anti-soliton) approaching the defect is moving slowly with $e^\theta < \sigma$.

\p Using the defect to remove a soliton or anti-soliton is not by
itself enough to reproduce the full variety of logic gates. For
example, if the first soliton to reach the gate is removed, the
second must be travelling slower and will be inverted. This is
illustrated in Table 1 where the first soliton to arrive at the defect is
labelled $a_1$ and the second $a_2$.
On the other hand, if the second arrival is to be removed the
first will pass the defect delayed yet retain its character.
Neither of these is especially useful but together they exhaust the
possibilities with a passive defect.

\p
\begin{center}
\begin{tabular}{|c|c||c|}
\hline
  $a_2$ & $a_1$ & $a_2*a_1$ \\ \hline
  1 & 1 & 0  \\
  1 & 0 & 0  \\
  0 & 1 & 1  \\
  0 & 0 & 1  \\
  \hline
\end{tabular}
\qquad\qquad
\begin{tabular}{|c|c||c|}
\hline
  $a_2$ & $a_1$ &\small $a_2$ \textsc{xor} $a_1$ \\ \hline
  1 & 1 & 0  \\
  1 & 0 & 1  \\
  0 & 1 & 1  \\
  0 & 0 & 0  \\
  \hline
\end{tabular}
\end{center}
\hspace{5.0cm} Table 1 \hspace{3.5cm} Table 2

\phantom{**}

\begin{center}
\begin{tabular}{|c|c|c||c|c|c|}
  \hline
  $a_3$ & $a_2$ & $a_1$ & \small $a_2$ \textsc{nand} $a_1$ & $a_2$ & $a_1$ \\
  \hline
  1 & 1 & 1 & 0 & 1 & 1 \\
  1& 1 & 0 & 1 & 1 & 0 \\
  1 & 0 & 1 & 1 & 0 & 1 \\
  1 & 0 & 0 & 1 & 0 & 0 \\
  \hline
\end{tabular}
\end{center}

\hspace{7.2cm} Table 3

\phantom{**}

\p
On the other hand, if it is supposed there is feedback, which allows the passage of a
soliton (but not an anti-soliton) to initiate a signal instructing the
defect `controller' to  raise the defect parameter, then the possibilities become
more interesting. For example, it would be possible to arrange the first arrival
to be removed and then raise the defect parameter if it is a soliton, but merely
to be removed if it is an anti-soliton leaving the defect parameter unchanged.
Under such circumstances, the second arrival will be inverted if the first arrival
is a soliton but not if it is an anti-soliton. This allows, in the same notation
as that used in Table 1, an XOR gate to
be represented (Table 2).

\p However, even with this device it is clear the output will
always have an even number of `1's and therefore both NOR and NAND (from
which all other standard two-bits-in-one-bit-out gates can be
constructed) are unobtainable. Besides, a careful tuning of the
defect parameter is required to remove a soliton and it might be
better to allow both solitons to pass, adjusting the defect
parameter each time there is a passing soliton, but not for a passing anti-soliton.
With two bits this will just reproduce the XOR gate already described.

\p
On the other hand, consider a triple of approaching solitons/anti-solitons. Each
will be affected by the defect independently of the others
(checked as before in the case of two using Hirota's explicit
solution), and arrange for each passing soliton (but not anti-soliton) to increase the
defect parameter by the same amount $\delta$ with
$$e^{\theta_1}>\sigma,\qquad
e^{\theta_2}>\sigma+\delta,\qquad
\sigma+2\delta>e^{\theta_3}>\sigma+\delta.$$
The three
arriving solitons/anti-solitons have rapidities $\theta_1> \theta_2
>\theta_3$.
With this
arrangement, the first pair of solitons/anti-solitons pass the defect, delayed
but without inversion, but the third is inverted provided the
first two were both solitons, and not otherwise. If the third is
always a soliton this provides a version of the NAND gate for the
first two (Table 3). In fact, using all possible triples in and
out with the rules described above gives a representation of the
Toffoli three-bit gate \cite{Toffoli80}

\section{Discussion}

As stated in the introduction, it is not yet clear if there is a set
of physical circumstances permitting the type of integrable defect
whose properties have been discussed. The search for such a system continues.
It would also be interesting to investigate further the quantum aspects of
field theories with defects (developing ideas originally pioneered by Delfino
et al. \cite{Delfino94}). In the past, the quantum sine-Gordon
model has been investigated (see, for example \cite{Konik97}), mostly
from an algebraic point of view, but it is not yet clear all the properties
outlined above (and those given in references \cite{bczlandau, bczjhep})
have been properly taken into account. In particular, no mention has been made
previously concerning a defect's ability to change topological charge
by $\pm 1$,  although it is clear the transmission matrix
discovered by Konik and LeClair does allow transitions between solitons and
anti-solitons changing topological charge by $\pm 2$.
Perhaps the transmission matrix constructed by them will need to be generalised,
or perhaps the capacity of a defect
to remove a soliton/anti-soliton does not survive quantisation; it is, after all,
a delicate matter since the incoming soliton rapidity needs to be precisely
matched to the defect parameter, yet quantum effects are rarely so
sharply tuned.

\p
Some time ago, Baseilhac and Delius discovered dynamical boundary conditions for
 the sine-Gordon model restricted to a half-line \cite{Baseilhac01}. Seeking the
analogues of these in association with a defect may prove profitable.

\p
One might also wonder
about the possibility of finding representations of two-qubit gates or
 a three-qubit gate, such as
the generalisation of the Toffoli gate due to Deutsch \cite{Deutsch89},
using solitons within an integrable quantum field theory.

\section{Acknowledgements}

One of us (CZ) thanks the  University of
York for a Studentship. Both of us have both benefited from
conversations with Peter Bowcock and other contacts within EUCLID - a European Commission
funded Research Training Network, contract number HPRN-CT-2002-00325.

\end{document}